\documentclass[10pt,letterpaper,conference]{IEEEtran}
\IEEEoverridecommandlockouts
\usepackage{cite}
\usepackage{amsmath,amssymb,amsfonts}
\usepackage{algorithmic}
\usepackage{graphicx}
\usepackage{textcomp}
\usepackage{xcolor}

\usepackage{color,soul}

\def\BibTeX{{\rm B\kern-.05em{\sc i\kern-.025em b}\kern-.08em
    T\kern-.1667em\lower.7ex\hbox{E}\kern-.125emX}}

\begin{document}

\title{Fully parallel implementation of \\ digital memcomputing on FPGA
\thanks{This work was supported by the NSF grant No. ECCS-2229880. Some of our results were obtained using the Anvil cluster at Purdue Rosen Center for Advanced Computing through allocation CIS240196 from the Advanced Cyberinfrastructure Coordination Ecosystem: Services \& Support (ACCESS) program, which is supported by NSF grants 2138259, 2138286, 2138307, 2137603, and 2138296.}
}

%\author{\IEEEauthorblockN{Dyk Chung Nguyen}
%\IEEEauthorblockA{\textit{Department of Physics and Astronomy} \\
%\textit{University of South Carolina}\\
%Columbia, South Carolina 29208, USA \\
%\and
%\IEEEauthorblockN{Yuanhang Zhang}
%\IEEEauthorblockA{\textit{dept. name of organization (of %Aff.)} \\
%\and
%\and
%\IEEEauthorblockA{\textit{dept. name of organization (of %Aff.)} \\

\author{\IEEEauthorblockN{Dyk Chung Nguyen
 and Yuriy V. Pershin
}
\IEEEauthorblockA{%\IEEEauthorrefmark{1}
Department of Physics and Astronomy, University of South Carolina, Columbia, South Carolina 29208, USA}
%\IEEEauthorblockA{\IEEEauthorrefmark{2}Department of Physics, University of California, San Diego, La Jolla, CA, 92093-0319, USA}
%\IEEEauthorblockA{\IEEEauthorrefmark{3}
Email: dykchung@email.sc.edu; pershin@physics.sc.edu}

\maketitle

\begin{abstract}
We present a fully parallel digital memcomputing solver implemented on a field-programmable gate array (FPGA) board. For this purpose, we have designed an FPGA code that solves the ordinary differential equations associated with digital memcomputing in parallel. A feature of the code is the use of only integer-type variables and integer constants to enhance optimization. Consequently, each integration step in our solver is executed in 96~ns. This method was utilized for difficult instances of the Boolean satisfiability (SAT) problem close to a phase transition, involving up to about 150 variables. Our results demonstrate that the parallel implementation reduces the scaling exponent by about 1 compared to a sequential C++ code on a standard computer. Additionally, compared to C++ code, we observed a time-to-solution advantage of about three orders of magnitude. Given the limitations of FPGA resources, the current implementation of digital memcomputing will be especially useful for solving compact but challenging problems.
\end{abstract}

\section{Introduction}

Digital memcomputing~\cite{Traversa17a,Sean3SAT,MemComputingbook} uses dynamical systems for computation. A dynamical system is characterized by a collection of ordinary differential equations (ODEs) that dictate the evolution of state variables. The structure of memcomputing ODEs ensures that, starting from any initial condition, the system will reach (after a suitable period) a stable state (attractor) that represents the solution to the problem.
Initially, the positions of such attractors are unknown, and the goal is to discover them via the continuous dynamics of the system. The effectiveness of memcomputing solvers is influenced by the approaches and strategies employed to solve their ODEs. More details on digital memcomputing can be found in a recent book~\cite{MemComputingbook}.

Although memcomputing ODEs can be solved using conventional computers, it is of interest to develop a hardware implementation that realizes the massively parallel dynamics of system variables~\cite{diventra13a}. %This approach can harness the inherent computational power of specialized hardware to significantly accelerate the processing speed and efficiency. 
On a smaller scale, the construction of digital memcomputing circuits is possible (but unpractical) with  off-the-shelf electronic  components~\cite{pershin23a,mem_analog_23a}. The present study focuses on the use of FPGAs as a feasible alternative to sequential software simulations. 

In the past, both FPGA and GPU (graphic processing unit) accelerators were employed to speedup various solvers, including those that rely on the numerical integration of ODEs. For example, in our recent work~\cite{chung23}, we employed a small-size  FPGA for numerical integration of memcomputing equations. However, the constraints of the small-size FPGA~\cite{chung23} only allowed for the development of a moderately effective solver through partial parallelization. Furthermore, Moln\'{a}r {\it et al.}~\cite{molnar2020accelerating} used GPUs to enhance the performance of a continuous-time analog SAT solver, achieving improvements of up to two orders of magnitude over traditional CPU-based methods. Sohanghpurwala {\it et al.}~\cite{sohanghpurwala2017hardware} authored a review on hardware-accelerated SAT solvers. For an extensive examination of different hardware accelerators, refer to Ref.~\cite{peccerillo2022survey}. 

In contrast to our earlier efforts documented in~\cite{chung23}, where
it was necessary to split each integration step into smaller steps to process each clause sequentially due to resource constraints, in the current work we use a larger FPGA (specified below), enabling a comprehensive parallel execution of digital memcomputing dynamics for problems involving a considerable number of variables.

\begin{figure*}[t]
\centering
   (a) \includegraphics[width=0.45\textwidth]{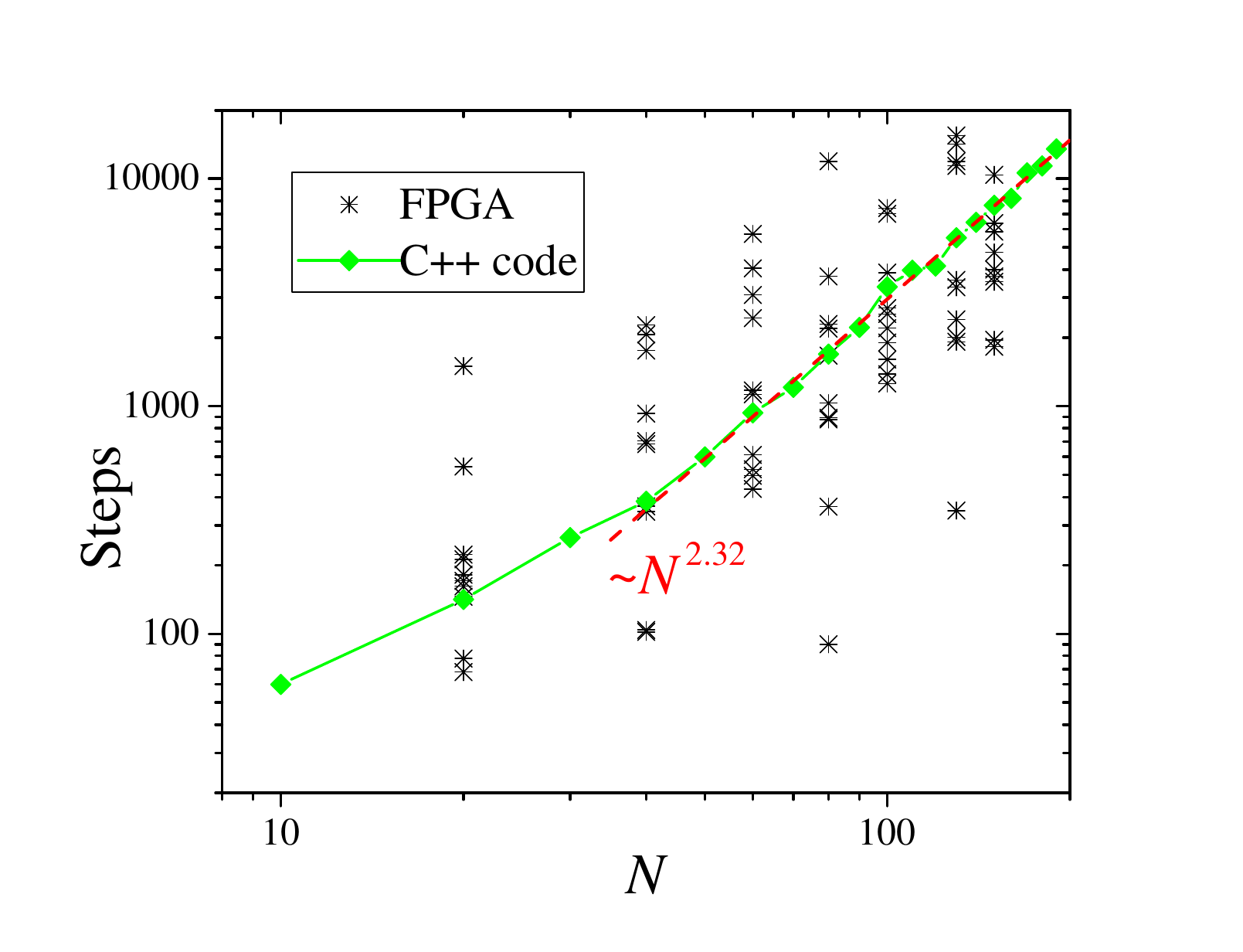}
    (b) \includegraphics[width=0.45\textwidth]{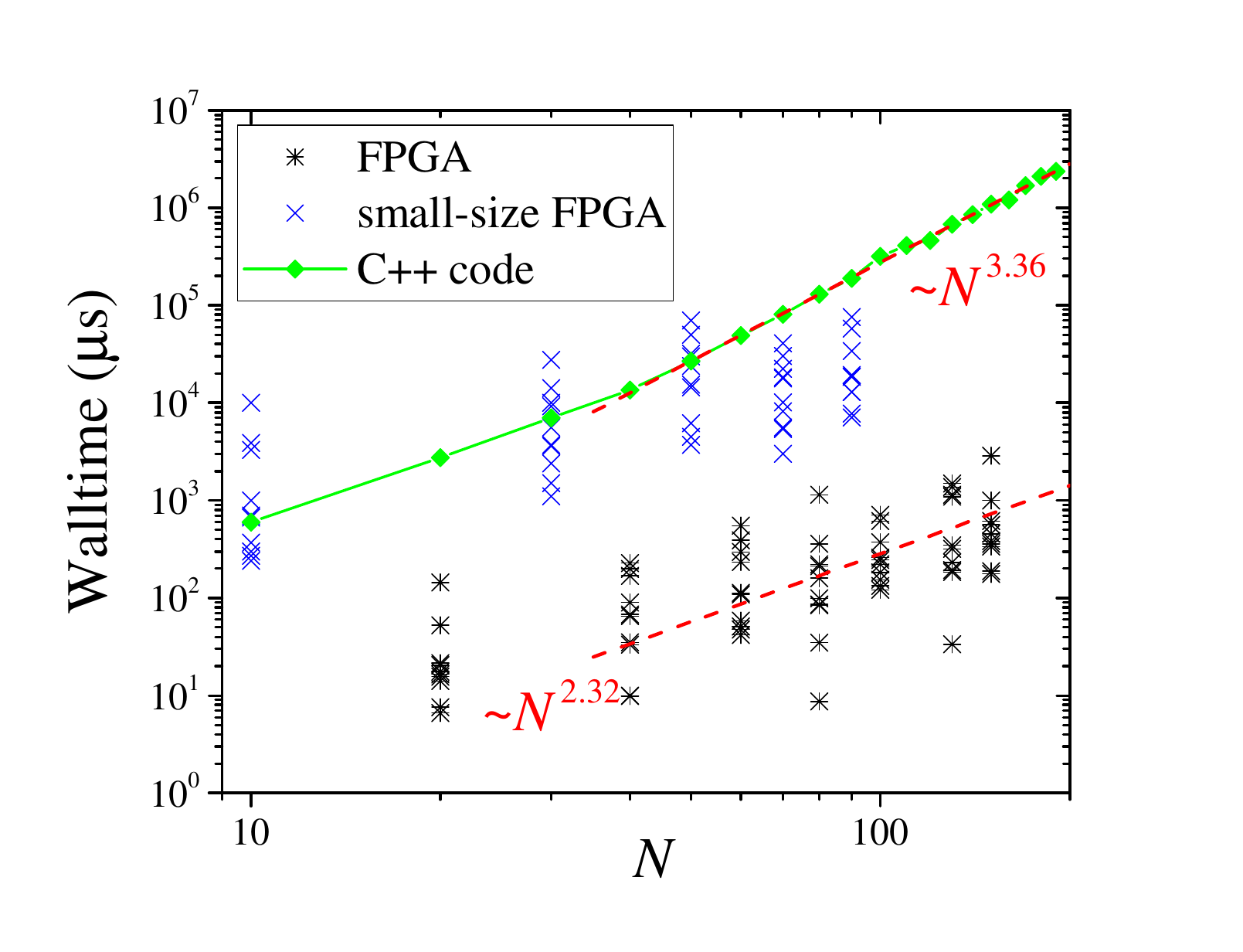}
    \caption{Vivado simulations of the fully parallel solution of Barthel $M/N=4.3$ instances on FPGA. (a) Steps to solution on FPGA (shown for ten instances of each selected problem size) and median number of steps to solution found using the C++ code. (b) Time to solution on FPGA (this work), small-size FPGA (from~\cite{chung23}), and C++ code.}
    \label{fig:1}
\end{figure*}

The structure of this conference paper is as follows. In Sec.~\ref{sec:2}, we introduce the digital memcomputing equations forming the basis for this work. Following this, in Sec.~\ref{sec:3}, we provide some technical information on the methods, including the generation of 3-SAT problem instances and the information on our FPGA and C++ code. Our results are detailed in Sec.~\ref{sec:4}.  The paper is concluded with a discussion and some concluding remarks in Sec.~\ref{sec:5}. 

\section{memcomputing equations for 3-SAT} \label{sec:2}

In this study, we use digital memcomputing equations from Ref.~\cite{Sean3SAT}, which were specifically developed to address a particular type of Boolean satisfiability problem, known as 3-SAT. The objective in 3-SAT is to determine a variable assignment that satisfies all clauses. Each clause is a disjunction of three literals; therefore, the entire Boolean formula is satisfied when at least one literal is TRUE. Here, a literal refers to either a Boolean variable or its negation. 

In short, the present work is based on the following equations~\cite{Sean3SAT}:

\begin{eqnarray}
\dot{v}_n&=&\sum\limits_mx_{l,m}x_{s,m}G_{n,m}(v_n,v_j,v_k)+\left( 1+\zeta x_{l,m}\right)\cdot \nonumber \\
& & \left(1-x_{s,m}\right) R_{n,m}(v_n,v_m,v_k), \label{eq:1}\\
\dot{x}_{s,m}&=&\beta \left( x_{s,m} +\epsilon \right)\left( C_m(v_i,v_j,v_k)-\gamma\right), \label{eq:2}\\
\dot{x}_{l,m}&=&\alpha \left( C_m(v_i,v_j,v_k)-\delta\right), \label{eq:3}\\
G_{n,m}&=&\frac{1}{2}q_{n,m}\text{min}\left[\left( 1-q_{j,m}v_j\right), \left( 1-q_{k,m}v_k\right)\right], \label{eq:4}\\
R_{n,m}&=&\begin{cases}
    \frac{1}{2}\left(q_{n,m}-v_n \right), \\
    \hspace{1cm} \text{if } C_m(v_n,v_j,v_k)=\frac{1}{2}\left(1-q_{n,m}v_n \right),\;\;\\
    0, \hspace{7mm} \text{otherwise}.
  \end{cases}
\end{eqnarray}
Here, $v_n$ are continuous versions of the Boolean variables ($n=1,\ldots,N$),  $x_{s,m}$ and $x_{l,m}$ are the memory variables ($m=1,\ldots,M$), where $N$ is the number of variables, $M$ is the number of clauses, $q_{j,m}=1$ if the $j$-th variable enters $m$-th clause, $q_{j,m}=-1$ if the negation of the $j$-th variable enters $m$-th clause. Moreover, $\alpha$, $\beta$, $\gamma$, $\delta$, $\epsilon$ and $\zeta$ are constants~\cite{Sean3SAT}. 

It is important to note that in the equations above, each 3-SAT variable is denoted by a continuous variable $v_n$, with each clause linked to two types of memory variables: short (s) and long (l). Furthermore, the range of $v_n$ is limited to $[-1,1]$, the range of $x_{s,m}$ to $[0, 1]$, and $x_{l,m}$ to $[1,10^4M]$. The Boolean value of $v_n$ is determined by its sign, with negative being 0 and positive being 1. 

Additionally, the clause function $C_m(v_i,v_j,v_k)$ is specified as
\begin{align}
& C_m(v_i,v_j,v_k)= \nonumber \\     
 &  \hspace{3mm}  \frac{1}{2}\text{min}\left[\left(1-q_{i,m}v_i \right),\left(1-q_{j,m}v_j \right),\left(1-q_{k,m}v_k \right)\right].\label{eq:6}
\end{align}
This expression represents the variable that best satisfies the clause $m$. For further information on the model and its derivation, see Refs.~\cite{Sean3SAT,MemComputingbook}.

\section{Methods} \label{sec:3}

\subsection{Generation of SAT instances}

Random SAT instances were generated using a code based on Ref.~\cite{barthel2002hiding}, henceforth termed Barthel instances. The specific variant of Barthel instances employed in our study is characterized by $M/N=4.3$. This ratio of $M/N$ closely approaches the phase transition~\cite{crawford1996experimental}, indicating that the instances are notably difficult to resolve. However, their construction method~\cite{barthel2002hiding} guarantees that they are solvable.

\subsection{FPGA implementation} 

In this study, we used an EK-U1-VCU118-G board equipped with a VU9P device
(Virtex UltraScale+ XCVU9P-L2FLGA2104E FPGA). This device contains 1182240 lookup tables (LUTs) and 6840 digital signal processors (DSPs). Verilog~\cite{palnitkar2003verilog,thomas2008verilog} was chosen for programming the FPGA due to its syntax, which is similar to C, which facilitates straightforward and efficient coding.

Eqs.~(\ref{eq:1})-(\ref{eq:3}) were solved using the forward Euler method. To enhance computational efficiency, integer data types were exclusively utilized. For this purpose, Eqs.~(\ref{eq:1})-(\ref{eq:3}) were scaled by a factor of $2^{14}$, and new  variables (scaled by this factor) were defined. In terms of these new variables, Eqs.~(\ref{eq:1})-(\ref{eq:3}) are rewritten as
\begin{eqnarray}
    \dot{V}_n & = &\sum_{m} \frac{X_{l,m}X_{s,m}}{2^{14}}  \frac{G'_{n,m}(V_n,V_j,V_k)}{2^{14}} + \\
    & & \frac{(2^{14} + \zeta X_{l,m})(2^{14} - X_{s,m})}{2^{14}} \frac{R'_{n,m}(V_n,V_j,V_k)}{2^{14}}, \nonumber \\
    \dot{X}_{s,m} &=& \frac{ \beta (X_{s,m} + 2^{14} \epsilon)(C'_{m}(V_n,V_j,V_k) - 2^{14} \gamma)} {2^{14}}, \\
 \textnormal{and}\;\;\;   & &  \nonumber\\
    \dot{X}_{l,m} &=& \alpha (C'_m(V_n,V_j,V_k) - 2^{14} \delta).
\end{eqnarray}
Here, $G'_{n,m}(\dotso)$, $R'_{n,m}(\dotso)$, and $C'_m(\dotso)$ are appropriately scaled versions of the functions defined in Eqs.~(\ref{eq:4})-(\ref{eq:6}).

Additionally, since multiplication and division by powers of two can be efficiently performed using bit shifts, the constants in the model were chosen to be powers of two whenever suitable. In particular, we set $\alpha = 4$, $\beta = 16$, $\gamma = 2^{-2}$, $\delta = 819\cdot2^{-14} \approx 0.05$, $\epsilon = 2^{-10}$, and $\zeta = 2^{-10}$. The integration time step, $\Delta t$, was set to $0.0625 = 2^{-4}$.

The UART interface facilitated the transmission of the problem solution from the FPGA board to the PC. The actual time required to solve the problem was independently recorded using a data acquisition module that was connected to an output pin on the FPGA board. During the calculation, the pin state was maintained at a low level (0) and switched to a high level (1) at other times.

To expedite the process, some of the results presented below were derived from simulations in Vivado, without the implementation phase. We note that the number of steps required to reach a solution remains consistent between the Vivado simulations and the real FPGA implementation of the same problem.

\subsection{C++ code}

The C++ programming language was used to implement the dynamic equations Eqs.~(\ref{eq:1})-(\ref{eq:3}). Numerical integration
was performed using the forward Euler method with exactly the same parameter values as in the FPGA. 
For each size of the problem, $10^3$ simulations were performed to extract the median time to solution. These simulations were carried out on the Anvil cluster at Purdue University, which was accessed through the NSF ACCESS program~\cite{boerner2023access}.

\begin{figure}[t]
\centering
    \includegraphics[width=0.45\textwidth]{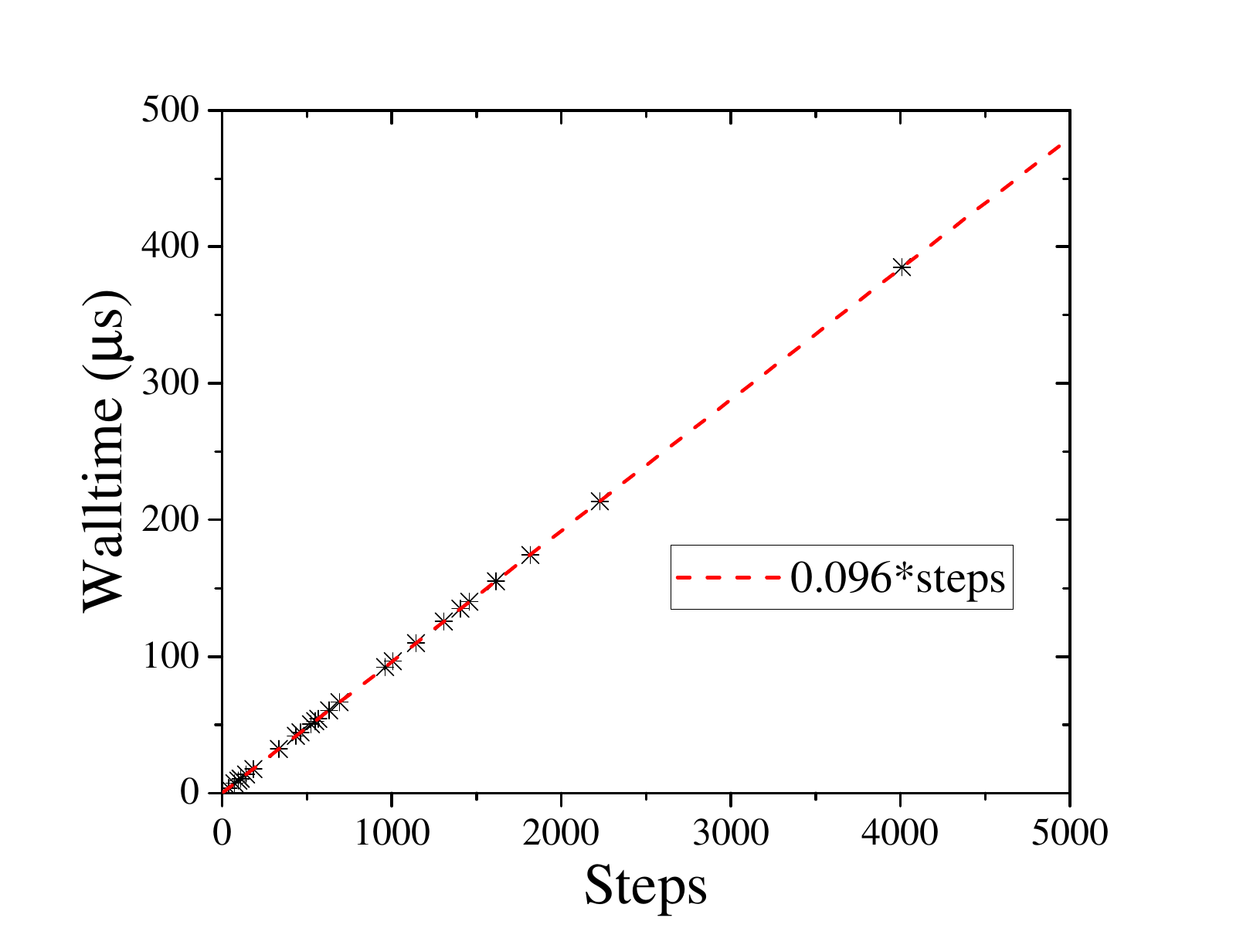} \\
    \caption{Experimental data showing the relationship between the time to solution and number of integration steps for various $M/N=4.3$ problems. The relationship is evidently linear.}
    \label{fig:2}
\end{figure}

\section{Results} \label{sec:4}

Our approach has enabled obtaining a 96~ns duration for each integration step, regardless of the size of the problem. We emphasize that within this time the solution of all differential equations in the model advances by one step in parallel.

Fig.~\ref{fig:1}(a) shows the number of steps to solution for problems of various sizes on FPGA ($N=20, 40, 60, 80, 100, 130, 150$).
Each problem size presented in this figure was tested with ten individual Barthel instances. C++ simulations has allowed an accurate determination of the median number of steps to solution, corroborating the FPGA findings. Analysis of the C++ data reveals that the number of steps to solution increases polynomially, characterized by a scaling exponent of $a=2.32\pm 0.04$.

Fig.~\ref{fig:1}(b) illustrates the real time to solution for various methods, including our earlier results for the small-size FPGA~\cite{chung23}. The data points for FPGA have been obtained by multiplying the number of steps, Fig.~\ref{fig:1}(a), by the duration of step, 96~ns. Our results indicate that the data points found with the C++ code align with a polynomial model, characterized by a scaling exponent of $b=3.36\pm 0.04$. 

Additional instances of the 3-SAT problem with a ratio of $M/N=4.3$ were implemented and executed on FPGA, and their characteristics were documented. In particular, Fig.~\ref{fig:2} shows that the time to solution scales linearly with the number of steps to solution, as expected. To obtain Fig.~\ref{fig:2}, we used the data acquisition module to measure the actual time to solution, while the number of steps was taken from the Vivado simulations.

Fig.~\ref{fig:3} shows that the number of LUTs scales linearly with the problem size starting at $N=40$. We note that the number of DSPs, $N_{DSP}$, also scales linearly with $N$ as $N_{DSP}=43\cdot N$. These observations are useful to project the current implementation to larger state-of-the-art FPGA devices.

\section{Discussion and conclusion} \label{sec:5}

\begin{figure}[t]
\centering
    \includegraphics[width=0.45\textwidth]{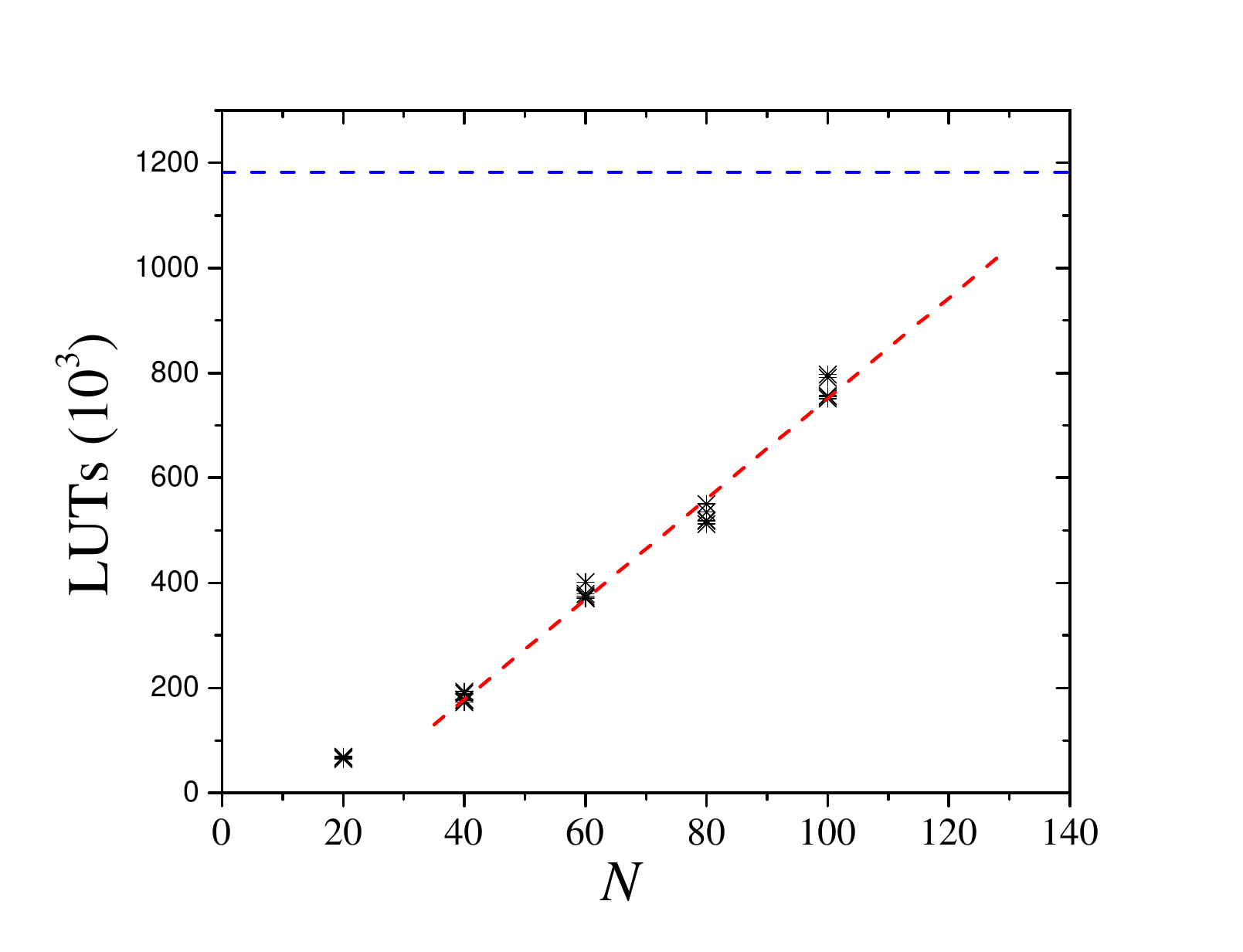}
    \caption{Utilization of LUTs depending on the problem size. The dashed horizontal line denotes the maximum number of LUTs in our VCU118 evaluation board. The fitting curve is $-204557+9559\cdot N$.}
    \label{fig:3}
\end{figure}

In this conference paper, we have presented the first significant implementation of  digital memcomputing dynamics in a fully parallel manner~\footnote{In our earlier work~\cite{chung23}, a fully parallel implementation was reported for a very small problem involving 3 variables and 6 clauses}. The effectiveness of our FPGA code is particularly due to the exclusive use of integer-type variables and integer numbers and the application of shifts to perform multiplication and division operations (whenever possible). Our strategies have resulted in a single-step integration of hundreds of ODEs within a short time interval of 96~nanoseconds.

It is remarkable that digital memcomputing machines, which are defined by ordinary differential equations, maintain their robustness when transitioning to the discrete domain. We highlight that the C++ code utilized in this research was written using floating-point variables. It is a striking fact that the FPGA data closely follow the trends observed in the C++ code simulations, as illustrated in Fig.~\ref{fig:1}(a). 

Our findings indicate that the parallel implementation decreases the time to solution in a polynomial manner. This observation has been confirmed through the application of~Eqs.~(\ref{eq:1})-(\ref{eq:3}) across different types of problems, including Barthel $M/N=7$ and XORSAT problems. Although the polynomial reduction is significant and potentially useful in practical cases, it does not convert the exponential complexity~\cite{kowalsky20223,pershin2024acceleration} of the approach based on Eqs.~(\ref{eq:1})-(\ref{eq:3}) into a polynomial one.
It remains unresolved whether a specific set of parameters can enable Eqs.~(\ref{eq:1})-(\ref{eq:3}) to address XORSAT problems in polynomial time.

Regarding the scalability of the hardware resources shown in Fig.~\ref{fig:3}, the VCU118 FPGA board utilized in this research is capable of addressing 3-SAT $N/M=4.3$ problems with up to about $150$ variables. It should be noted that modern state-of-the-art FPGA devices are powered by an order of magnitude larger number of LUTs. Consequently, we expect that such state-of-the-art FPGA devices could handle an order of magnitude larger problems implemented in a similar fashion.

In conclusion, through the use of an FPGA board, we have achieved, for the first time, a fully parallel implementation of digital memcomputing dynamics on a significant scale. We have shown a notable enhancement over the sequential integration of ordinary differential equations, particularly in achieving a polynomial reduction in the time to solution. In a more general context, this study highlights the potential of FPGAs in addressing complex optimization problems.

\section*{Acknowledgement}

The authors express their gratitude to Yuan-Hang Zhang and Massimiliano Di Ventra for many interesting discussions.

\newpage

\bibliographystyle{IEEEtran}
\bibliography{biblio}

\end{document}